%
%
%
%

\input harvmac.tex


\noblackbox

\def\a#1{{A_{#1}}}

\def\ad#1{{A_{#1}^\dagger}}

\def\ada{{a^\dagger}}

\def\bt{{\theta^T}}

\def\tir{{\tilde r}}

\def\bZ{{\bf Z}}


\def\np#1#2#3{Nucl. Phys. {\bf B#1} (#2) #3}
\def\pl#1#2#3{Phys. Lett. {\bf #1B} (#2) #3}

\lref\bfss{T. Banks, W. Fischler, S. H. Shenker and L. Susskind, ``M
Theory As A Matrix Model: A Conjecture'', hep-th/9610043.}
\lref\green{M. B. Green, ``Pointlike States for type IIB
Superstrings'', \pl{329}{1994}{435}, hep-th/9403040.}
\lref\bankssus{T. Banks and L. Susskind, ``Brane-antibrane forces'',
hep-th/9511194.}
\lref\duffstelle{M. J. Duff and K. S. Stelle, ``Multimembrane solutions of
$D=11$ supergravity'', \pl{253}{1991}{113}.}
\lref\duff{M. J. Duff, ``M Theory (the Theory Formerly Known as
Strings)'', hep-th/9608117.}
\lref\asy{O. Aharony, J. Sonnenschein and S. Yankielowicz,
``Interactions of Strings and D-branes from M Theory'',
\np{474}{1996}{309}, hep-th/9603009.}
\lref\dbound{E. Witten, ``Bound States of Strings and P-branes'',
\np{460}{1996}{335}, hep-th/9510135.}
\lref\dbraneone{U. H. Danielsson, G. Ferretti and B. Sundborg,
``D-particle Dynamics and Bound States'', hep-th/9603081.}
\lref\dbranetwo{D. Kabat and P. Pouliot, ``A Comment on Zero-Brane Quantum
Mechanics'', hep-th/9603127.}
\lref\dbranethree{
M. R. Douglas, D. Kabat, P. Pouliot and S. H. Shenker, ``D-branes and
Short Distances in String Theory'', hep-th/9608024.}
\lref\memb{B. de Wit, J. Hoppe and H. Nicolai, ``On the Quantum
Mechanics of Supermembranes'', \np{305}{1988}{545} \semi
B. de Wit, M. Luscher and H. Nicolai, ``The Supermembrane is
Unstable'', \np{320}{1989}{135}.}
\lref\susskind{L. Susskind, ``T Duality in M(atrix) Theory and S
Duality in Field Theory'', hep-th/9611164.}
\lref\ori{O. J. Ganor, S. Ramgoolam and W. Taylor IV, ``Branes, Fluxes
and Duality in M(atrix) Theory'', hep-th/9611202.}
\lref\me{O. Aharony, ``String Theory Dualities from M Theory'',
\np{476}{1996}{470}, hep-th/9604103.}
\lref\berdoug{M. Berkooz and M. R. Douglas, ``Five-branes in M(atrix)
Theory'', hep-th/9610236.}
\lref\schwarz{J. H. Schwarz, ``Lectures on Superstring and M Theory
Dualities'', hep-th/9607201.}
\lref\bound{J. G. Russo and A. A. Tseytlin, ``Waves, Boosted Branes
and BPS States in M Theory'', hep-th/9611047 \semi
J. C. Breckenridge, G. Michaud and R. C. Myers, ``More D-brane Bound
States'', hep-th/9611174.}
\lref\bachas{C. Bachas, ``D-brane Dynamics'', hep-th/9511043.}

%
%

\Title{RU-96-106, hep-th/9611215}
{\vbox{\centerline{Membrane Dynamics in M(atrix) Theory}}}
\centerline{Ofer Aharony and Micha Berkooz
\footnote{$^\star$}
{oferah, berkooz@physics.rutgers.edu 
}}
\bigskip\centerline{\it Department of Physics and Astronomy}
\centerline{\it Rutgers University}
\centerline{\it Piscataway, NJ 08855-0849}

\vskip .3in
We analyze some of the kinematical and dynamical
properties of flat infinite membrane solutions in the conjectured M
theory proposed by Banks, Fischler, Shenker and Susskind. In
particular, we compute the long range potential between membranes and
anti-membranes, and between membranes and gravitons, and compare it
with the supergravity results. We also discuss membranes with finite
relative longitudinal velocities, providing some evidence for the
eleven dimensional Lorentz invariance of the theory.

\Date{11/96} 

%
\newsec{Introduction}

Recently, Banks, Fischler, Shenker and Susskind
have proposed \bfss\ a definition of eleven dimensional
M theory\foot{See \refs{\schwarz,\duff} for recent reviews of M theory.}
in the infinite momentum frame as a large $N$ limit of
maximally supersymmetric matrix quantum mechanics. From the string
theory point of view, this description arises as the theory governing
the short distance dynamics of D0-branes in type IIA string theory
\refs{\dbound,\dbraneone,\dbranetwo,\dbranethree}
(which goes over to M theory in the strong coupling
limit). Surprisingly, the same system arises also as a regularization
of eleven 
dimensional supermembrane theory in the lightcone frame \memb. This is
one of the hints that the proposal of Banks et al, that naively
describes only 0-branes, in fact includes all of the degrees of
freedom of M theory. Another strong hint for this is provided by the
recent papers \refs{\susskind,\ori}, which showed that when compactified
on a 3-torus to 8 dimensions, M(atrix) theory has an $SL(2,\bZ)$
invariance that corresponds to the $SL(2,\bZ)$ factor in the
$SL(3,\bZ)\times SL(2,\bZ)$ U duality group of 8 dimensional type II
string theory (the $SL(3,\bZ)$ factor is trivial from the M theory
point of view). Since this duality, together with the geometrical
symmetries of M theory compactifications, may be used to generate all
of the known string theory dualities \me, this suggests that M(atrix)
theory indeed reproduces all the dualities of M theory (assuming, of
course, that M(atrix) theory has eleven dimensional Lorentz
invariance). Since this
duality exchanges a membrane with a fivebrane wrapped around the
3-torus, it also suggests that M(atrix) theory includes also
fivebranes. It is not yet
known how to describe in M(atrix) theory fivebranes that are not
wrapped around the longitudinal direction, but a consistent description of
wrapped fivebranes was given in \berdoug.

In this paper we analyze in detail some properties of the membrane
configurations of M(atrix) theory, and check that they correspond to
our expectations from membranes in M theory. We begin in section 2 by
reviewing the description of membranes in M(atrix) theory, and
analyzing their kinematical properties. 
In section 3 we
compute the long range potential between a membrane and an anti-membrane,
by computing the zero point energy of the corresponding configuration
in the quantum mechanics. We reproduce the expected $1/r^5$ behavior
of the potential, up to numerical constants.
At short distances, we find that a tachyonic
mode develops when the distance between the membrane and the
anti-membrane is of the order of the string scale (which goes to zero
in the eleven dimensional limit). In section 4 we compute the long
range potential between membranes with a relative longitudinal velocity,
and show that it matches our expectations from M theory for a
$v^4/r^5$ behavior (up to numerical constants). This is a
direct check of eleven dimensional Lorentz invariance in M(atrix)
theory. In section 5 we
compute the long range potential between a membrane and a 0-brane (or a
graviton). We end in section 6 with a summary of our results and a
discussion of some open questions.

\newsec{Infinite membranes in M(atrix) theory}

As discussed in \bfss, the large $N$ limit of the matrix quantum
mechanics seems to contain membrane configurations which break half of
the supersymmetry. In this section we
discuss some aspects of the kinematics of infinite membranes, before
performing calculations with these membranes in the next sections.

The Hamiltonian of M(atrix) theory \bfss\ is given by
\eqn\hamilton{H = R \tr\{{{\Pi_i \Pi_i}\over 2} + {1\over 4}[Y^i,Y^j]^2
+ \theta^T \gamma_i [\theta,Y^i]\}.}
Before beginning, we need to be more precise about what is the meaning
of the  Hamiltonian \hamilton. Our conventions\foot{Our conventions
for indices are the following. Indices in the standard frame run from
0 to 9 and 11, and the metric is $diag(-1,1,\cdots,1)$. Light-cone
coordinates are defined by $X^{\pm} = {1\over \sqrt{2}}(X^0 \pm
X^{11})$ and $\gamma^{\pm} = {1\over \sqrt{2}}(\gamma^0 \pm
\gamma^{11})$. Coordinates will generally have upper indices, and
momenta will have lower indices. $\gamma^{ij}$ is defined as
$\gamma^{ij} = {1\over 2}[\gamma^i,\gamma^j]$.} differ slightly from
those of \bfss.  The Hamiltonian \hamilton\ is $P_+$ (which equals
$P^-$), and, therefore, it equals $m^2 / 2P_-$ (when $P_i=0$). 
$P_-$ is quantized
to be $N / R$. These conventions are the natural ones for the light
cone frame, and they are the only ones that give consistent kinematics
for generic configurations\foot{They are different from the
conventions of \bfss\ where $P_{11}$ was taken to be $N/R$.}.

An infinite membrane spanning the coordinates $Y^1$ and $Y^2$ is
described by a configuration in which
$[Y^1,Y^2] = 2\pi i z^2$. Since the commutator has non-zero trace,
such configurations are obviously impossible for finite $N$.
We will
regard $Y^{1,2}$, for finite N, roughly\foot{We should caution the
reader that we have not made precise our definition of
the $N\to \infty$ limit in this regime. But, at this stage, we
are using the finite $N$ regulator only in a mild fashion, to cut off the
area of the membrane. We will return to this issue later.} 
as $Y^1=z\sqrt{N}q,
Y^2=z\sqrt{N}p$ where $p,q$ are defined in \bfss. The information we
need about the matrices $p$ and $q$, for our purposes, is that their spectrum
can be taken to go from $0$ to $2\pi$. This suggests that the membrane
extends from $0$ to $2\pi z\sqrt{N}$ in each direction, and, thus, its
area is $(2\pi)^2 z^2 N$. Plugging this
solution into the Hamiltonian one finds that $H = P_+ = R N (2\pi
z^2)^2 / 2 = m^2 / 2 P_-$, and, therefore, the mass of the membrane
is $m = 2\pi z^2N$. Comparing to the area, we see that the tension of the
membrane is $1 / 2\pi$. This is consistent with the calculation of
the membrane tension in the appendix of \bfss.

With these assumptions, we have completely characterized the
kinematics of the membrane, and any additional relation will test some of
our assumptions. In particular, we can check the assumption that
$P_+=H$, which in our case gives $P_+ / m=R\pi z^2$, by comparing
it to the value derived from the unbroken SUSYs. 
The SUSY transformation of the fermionic coordinates, for static
configurations,
may be written as \bfss
\eqn\susytrans{\delta \theta = {1\over 2} \gamma^+ ({R\over 2} 
[Y^i,Y^j] \gamma^{ij} + \gamma^-) \epsilon.}
Thus, the infinite membrane 
configuration breaks half of the supersymmetries, and we can
also use this relation to examine the kinematics of the membrane. The
membrane solution defined above preserves supersymmetry generators such that
$\gamma^+ (R 2\pi i z^2 \gamma^{12} + \gamma^-) \epsilon = 0$.
A supermembrane in the rest frame conserves only
supersymmetry generators $\epsilon$ such that $\gamma^0 \gamma^{12}
\epsilon = i \epsilon$, and in a boosted frame this becomes
\eqn\boostsusy{
{1\over m}(P_+ \gamma^+ + P_- \gamma^-) \gamma^{12} \epsilon = i \epsilon.}
Multiplying \boostsusy\ by $i \gamma^+ \gamma^-$, we find (using
$(\gamma^-)^2=0$ and $\{\gamma^+,\gamma^-\} = 2I$) that $\gamma^+
(2 i {P_+\over m} \gamma^{12} + \gamma^-) \epsilon = 0$, so we should
identify ${P_+ / m}=\pi z^2R$, which is consistent with what we
found before. 

For completeness, let us make the distinction between our infinite
membrane configurations and the ``finite''
membrane configurations more precise. 
By finite membranes we mean membranes defined by another natural
rescaling of $Y^{1,2}$ with N, in which we take $Y^1 = R_1 q$ and 
$Y^2 = R_2 p$, corresponding to $z = \sqrt{R_1
R_2 / N}$. From the spectrum of $p$ and $q$ we see that
this apparently corresponds to a membrane of area $(2\pi)^2 R_1 R_2$
and mass $2\pi R_1 R_2$ (which again gives a tension
${1 / 2\pi}$). This type of configuration has a
finite rest mass, finite area and infinite longitudinal momentum
density and, like the graviton configurations
discussed in \bfss, is one for which the infinite  momentum frame
Hamiltonian is the same as the light cone Hamiltonian (up to numerical
factors), which is an advantage. The disadvantage is that finite size
membranes are not BPS-saturated, which makes it more difficult to analyze
issues such as charges of the 3-form field.
We certainly do not expect to have
sensible configurations that look like an $R_1\times R_2$ bit
of membrane,
so we need work with configurations of closed
membranes, and this leads to much more difficult computations, which
have not yet been performed.
To summarize, the finite membranes have finite size and infinite
longitudinal momentum density whereas the infinite membrane has an
area that grows like $N$, giving a finite momentum density. 

For infinite membranes, unlike
gravitons, we can now go to the rest frame of the membrane by a
finite (N independent) boost. In light cone coordinates, boosts act as
$P_- \to \gamma P_-$ and $P_+ \to P_+ / \gamma$. In the rest frame
$P_+ = P_-$, so it is easy to see that the boost factor we need is
$\gamma = \sqrt{2} \pi R z^2$.
In the lightcone we can loosely think of the system as
satisfying (for any wave function $\phi$)
$\phi(X^+,X^-)=\phi(X^+,X^- +2\pi R)$. Boosting by $\gamma$ to
go to the rest frame of the membrane, we find
$\phi(t,x^{11})=\phi(t+{\sqrt{2} \pi R\over \gamma},
x^{11}+{\sqrt{2} \pi R\over \gamma})$.
Since we are discussing a configuration that is static in this frame, its
wave function satisfies $\phi(t,x^{11})=\phi(t,x^{11}+{\sqrt{2} \pi
R\over \gamma})$,
which determines the radius of the eleventh dimension in the rest frame
of the branes to be $R^{11}_{rest} = R / \sqrt{2} \gamma = 1 / 2\pi z^2$.
Since we are dealing with configurations that have a definite momentum
in the longitudinal direction, the wave functions of our membranes
will be spread out in this direction, and we will have to average over
it in our computations. Note that $R^{11}_{rest}$ that we found does
not depend on $R$, so the infinite membrane system does not become
decompactified as $R\rightarrow\infty$ 
(but rather when $z^2\rightarrow 0$), unlike the situation for finite
energy configurations. 

Configurations of several gravitons are described in the matrix theory
by block matrices, each of whose blocks corresponds to a single
graviton (and such that all of the blocks go to infinite size in the
large $N$ limit). In the same way we can describe multiple membrane
configurations, by taking several ``membrane blocks'' with different
values of the transverse coordinates. As long as the membranes are
parallel (and have the same value of $z$), 
such a configuration still breaks only half of the
supersymmetry.

\newsec{Membrane-anti-membrane dynamics}

Anti-0-branes are not present in the matrix model, since the infinite
boost we performed has turned them all into 0-branes. However,
``anti-membranes'' in M theory are just membranes with an opposite
orientation, and these have the same status as membranes in the
infinite momentum frame. We can easily discuss such an anti-membrane
by choosing a configuration with an opposite value of $[Y^1,Y^2]$,
since this is identified with the wrapping number of the membrane. For
instance, we can multiply $Y^2$ by $(-1)$ 
to turn a membrane configuration into
an anti-membrane
configuration. Note that the anti-membrane also breaks half of the
supersymmetry, but that if we take a configuration including both
membranes and anti-membranes, the supersymmetry is completely broken,
as expected. Configurations involving both gravitons and membranes
also break all of the supersymmetry, as they should. These are
discussed in section 5.

In this section we will discuss the dynamics of a system including one
membrane and one anti-membrane, with some distance $r$ between them
(which we will take to be along $Y^3$). Thus, our vacuum state will
correspond to a configuration of the form
\eqn\vacuum{Y^1_0=\pmatrix{Q_1 & 0 \cr 0 & Q_2 \cr};\qquad
Y^2_0 = \pmatrix{P_1 & 0 \cr 0 & -P_2 \cr};\qquad Y^3_0 = \pmatrix{0 &
0 \cr 0 & r\cr},}
where we will now normalize
$[Q_1,P_1]=[Q_2,P_2]=2\pi i z^2$, each block is of size $N\times N$,
and all the other $Y_0^i$ vanish. 

Eleven dimensional supergravity 
predicts that the potential between a membrane and an anti-membrane,
after averaging over the position in the longitudinal direction,
will go like ${1 / r^5}$. To compute this potential in M(atrix)
theory, we will integrate out the off-diagonal blocks in all of the
matrices, which are
the remains of strings stretching from the membrane to
the anti-membrane. The supersymmetry is completely broken in this
vacuum, since a different half of the SUSY is broken by each 
block\foot{We will
later return to some speculations regarding approximate SUSYs.}. Thus,
we would naively expect to have corrections that are much larger than
$1/r^5$. The non-trivial cancelation of all higher terms in the 
potential, such as the
${1 / r}$ and ${1 / r^3}$ terms, serves as a test of the
M(atrix) theory away from any obvious BPS limit. 

Another reason to suspect that we might run into problems, in addition
to the fact that the supersymmetry is completely broken, is the
following. In string theory, the large $r$ closed string tree level
potential between a membrane and an anti-membrane can not be computed in
terms of the low energy excitations of the field theory on the
D-branes. Instead, one needs to include the whole infinite tower of
states in the open string sector. But here, when going from the full 
0-brane description to the QM, we have explicitly dropped all these
states.

In section 3.1 we compute the masses of the relevant bosonic modes
in the presence of the membrane-anti-membrane pair, and set up the
general procedure for such computations. In section 3.2 
we do the same for the fermionic
modes. In section 3.3 we compute the long-range potential between the
membrane and the anti-membrane, and check that it agrees with the known
M theory result (up to numerical factors). In section 3.4 we give a
short discussion of the annihilation process.

\subsec{Bosonic off-diagonal quadratic terms}

The masses for the off-diagonal bosonic modes arise (at tree-level)
from the term $\sum_{1\le i < j \le 9}{R\over 2}[Y^i,Y^j]^2$ 
in the matrix model
Hamiltonian. We will expand this term around the vacuum \vacuum, and
keep only the quadratic terms in the off-diagonal modes, which we will
denote by $Y^\mu \sim \pmatrix{0 & A_\mu \cr A_\mu^\dagger & 0}$. We
will then use the Born-Oppenheimer approximation to calculate the
corrections to the energy of the configuration.

Expanding the potential in a straightforward
manner\foot{Maintaining operator ordering, to which we will
return momentarily.}, the terms we find for $\mu=4,5,\cdots,9$ are
\eqn\mtxpta{\eqalign{R [ & \tr(2Q_1A_\mu Q_2 A_\mu^\dagger -
A_\mu Q_2^2A_\mu^\dagger
-Q_1A_\mu A_\mu^\dagger Q_1)+ \cr
& \tr(-2P_1A_\mu P_2A_\mu^\dagger-A_\mu P_2^2A_\mu^\dagger
-P_1A_\mu A_\mu^\dagger P_1) - r^2\tr(A_\mu A_\mu^\dagger)].\cr}}

The term $[Y^1,Y^2]^2$ gives
\eqn\mtxptb{\eqalign{
R [\tr( & \a{1}\ad{2}[Q_1,P_1]-\a{2}\ad{1}[Q_1,P_1]-
\ad{1}\a{2}[Q_2,P_2]+\ad{2}\a{1}[Q_2,P_2])+ \cr
\tr( &
Q_1\a{2}\ad{1}P_1-\a{1}P_2\ad{1}P_1-P_1\a{1}\ad{1}P_1-\a{2}Q_2\ad{1}P_1+
\cr &
Q_1\a{2}Q_2\ad{2}-\a{1}P_2Q_2\ad{2}-P_1\a{1}Q_2\ad{2}-\a{2}Q_2^2\ad{2}-
\cr &
Q_1\a{2}\ad{2}Q_1+\a{1}P_2\ad{2}Q_1+P_1\a{1}\ad{2}Q_1+\a{2}Q_2\ad{2}Q_1+
\cr &
Q_1\a{2}P_2\ad{1}-\a{1}P_2^2\ad{1}-P_1\a{1}P_2\ad{1}-\a{2}Q_2P_2\ad{1})].
\cr}}

The $[Y^1,Y^3]^2$ term gives
\eqn\mtxptc{R\tr[(Q_1\a{3}+r\a{1}-\a{3}Q_2)(-\ad{1}r+Q_2\ad{3}-
\ad{3}Q_1)],}
and the $[Y^2,Y^3]^2$ term gives
\eqn\mtxptd{R\tr[(P_1\a{3}+r\a{2}+\a{3}P_2)(-\ad{2}r-P_2\ad{3}-\ad{3}P_1)].}

Since $[Q_i,P_i] = 2\pi i z^2$, we can represent them as differential
operators by $Q_i=\sigma_i$ and $P_i=-2\pi i z^2 \del_i$, and
convert the Hamiltonian into a differential operator (as in
\berdoug). In this language, the matrix $A^\mu$ becomes now a
function of $\sigma_1$ and $\sigma_2$, so we will end up with a 2+1
dimensional field theory (with space dependent couplings). 
Note that even though
each membrane has two coordinates, each
membrane contributes only one coordinate in this representation. 
Correspondingly, the
differential operator realization of the Hamiltonian here is quite different
from the realization we get when treating commutators as Poisson brackets, as
is done for the finite membrane \bfss. Thus, these two objects are very
different from the point of view of the light-cone formulation. One would
like to believe that taking the area of the finite membrane to
infinity will be
equivalent to infinitely boosting the infinite membrane, but a
concrete demonstration of this does not yet exist.  

There are two subtleties that need to be taken into account. The first
is that we will take the coordinates $\sigma_i$ to go from $-\infty$
to $\infty$. We will then use finite (large) $N$ to regulate the
area, assuming that for finite $N$ we can truncate the spectrum
consistently. In the best of all possible worlds, we would calculate
the exact spectrum for finite $N$, and then take $N$ to infinity. One way
of doing this could be to choose $\sigma_i$ to be in the line segment $[-\pi
z\sqrt{N},\pi z\sqrt{N}]$, and to also put a cut-off on the momentum of
the order $P_{max} = \pi z\sqrt{N}$, and to write the matrices in an
appropriate basis. 
However, we do not know how to perform an exact computation in this case.
Instead, what we do here is
take the cut-off on $\sigma$ and on the momentum to infinity,
obtaining the full real line with no restrictions on momenta. We
calculate the spectrum and the wave functions on the entire axis and
then regulate to finite $N$ by taking wave functions that are,
essentially, supported in the finite interval $[-\pi
z\sqrt{N},\pi z\sqrt{N}]$ (both in $\sigma$-space and in momentum
space). This yields the correct $z$ and $N$
dependence, but might easily introduce numerical factors that we
cannot determine at this level of precision.  

The second subtlety is related to the first one. For finite $N$ we know
exactly what is the class of matrices that we are integrating
out. However, once
we write them as functions it is not clear what class of functions is
the correct one. For example, are we allowed to take $A_\mu$ to be
singular (for example, to behave like a derivative operator) ?
This will make a difference in the formal manipulations that we will
do shortly, which include various integrations by parts, derivatives,
etc. Since at the end of the day we find that the Hamiltonian that
we want to diagonalize is an harmonic oscillator, we will take the usual
$L^2$ functions on the $\sigma_i$ lines, and we will use the
eigenfunctions of
the harmonic oscillator as a basis for the space of functions that we
allow. This point seems to be
more of a technical nuisance than a real issue.   

The rules we will use
for transforming a trace, such as the ones written above, into a field
theory expression, are 
\eqn\rules{\eqalign{\tr(\a{m}O_2\ad{n}O_1) & \rightarrow \eta(O_2)\int
A^*_nO_1O_2A_m \cr
\tr(\a{m}\ad{n}O_1O_1') & \rightarrow \int A_n^*O_1O_1'A_m \cr
\tr(\ad{m}\a{n}O_2O_2') & \rightarrow \eta(O_2)\eta(O_2')\int
A_m^* O_2' O_2 A_n, \cr}} 
where a 1(2) subscript denotes that this is
an operator of the form $\sigma_1$ ($\sigma_2$) or
$\partial_{\sigma_1}$ ($\del_{\sigma_2}$),
and $\eta(\partial_i)=-1,\eta(\sigma_i)=1$. 

Let us, for example, show how to derive the first relation for
$O_i=\partial_i$. We replace the trace by a sum over a complete set of
functions $H_n(\sigma)$ (such as eigenfunctions of the harmonic
oscillator), satisfying $\sum_n H_n(\sigma) H_n(\tau) =
\delta(\sigma-\tau)$. Every
summation over an index corresponding to the membrane is replaced by 
an integral
over $\sigma_1$, and every summation over an index corresponding to the
anti-membrane is replaced by an integral over $\sigma_2$. 
Using the explicit form
of the operators, we obtain
\eqn\rlprf{\eqalign{tr(A_mO_2A_n^\dagger O_1) &=\sum_n\int
d\sigma_1d\sigma_2d{\hat\sigma}_1H_n(\sigma_1)A_m(\sigma_1,\sigma_2)
\partial_{\sigma_2}A_n^*({\hat\sigma}_1,\sigma_2)\partial_{{\hat\sigma}_1}
H_n^*({\hat\sigma}_1)= \cr
&=-\sum_n\int d\sigma_1d\sigma_2d{\hat\sigma}_1
H_n(\sigma_1)H_n({\hat\sigma}_1)A_m(\sigma_1,\sigma_2)
\partial_{\sigma_2}\partial_{{\hat\sigma}_1}A_n^*({\hat\sigma}_1,\sigma_2)=
\cr
& =-\int d\sigma_1d\sigma_2
A_n^*(\sigma_1,\sigma_2)\partial_{\sigma_1}
\partial_{\sigma_2}A_m(\sigma_1,\sigma_2), \cr}}
where we have performed several integrations by parts that are well
defined on our class of functions.

Using these rules, and defining $P={1\over\sqrt{2}}(P_1-P_2)$ and
$Q={1\over\sqrt{2}}(Q_1-Q_2)$ (satisfying $[P,Q]=-2\pi iz^2$), the
relevant terms (equations \mtxpta-\mtxptd) become
\eqn\baca{-R\int A_\mu^*(2P^2+2Q^2+r^2)A_\mu,}
\eqn\bacb{\eqalign{-2R\int &
A_1^*P^2A_1+A_2^*Q^2A_2+A_3^*(P^2+Q^2)A_3+ \cr
& A_2^*(QP+2\pi iz^2)A_1+A_1^*(PQ-2\pi iz^2)A_2,}}
\eqn\bacc{R\int \sqrt{2}
(-rA_1^*QA_3-rA_3^*QA_1)-r^2A_1^*A_1}
and
\eqn\babd{R\int \sqrt{2} (-rA_2^*PA_3-rA_3^*PA_2)-r^2A_2^*A_2.}
The integrations were originally over both $\sigma_1$ and $\sigma_2$,
but since the Hamiltonian has no dependence on $\sigma_1+\sigma_2$, we
can turn them into integrations just over $\sigma_1-\sigma_2$, with
approximately an order $N$ degeneracy of all modes (corresponding to
multiplying the $A_i$ by an arbitrary function of
$\sigma_1+\sigma_2$). The terms we found involving $A_1,A_2$ and $A_3$
can be rewritten in the form 
\eqn\matfrm{-R\int \bigl(A_1^*,A_2^*,A_3^*\bigr)\Bigl(\matrix{
2P^2+r^2&-2(PQ-2\pi iz^2)&\sqrt{2}rQ\cr
-2(QP+2\pi iz^2)&2Q^2+r^2&\sqrt{2}rP\cr
\sqrt{2}rQ&\sqrt{2}rP&2(P^2+Q^2)}\Bigr)
\Bigl(\matrix{A_1\cr A_2\cr A_3}\Bigr).}

We can write these terms as
$-\int A^* M_1 A$ where
\eqn\nmatfrm{M_1 = 2R \pmatrix{
P^2+r^2/2&-(PQ-2\pi iz^2)&r Q/\sqrt{2}\cr
-(QP+2\pi iz^2)&Q^2+r^2/2&r P/\sqrt{2}\cr
r Q/\sqrt{2}&r P/\sqrt{2}&P^2+Q^2}.}

Conjugating by the unitary matrix
\eqn\unimat{U = \pmatrix{{1\over \sqrt{2}} & -{i\over \sqrt{2}} & 0
\cr -{1\over \sqrt{2}} & -{i\over \sqrt{2}} & 0 \cr 0 & 0 & 1 \cr}}
and defining $a = (Q + iP) / \sqrt{2(2\pi z^2)}$, $\ada = (Q - iP) /
\sqrt{2(2\pi) z^2}$ (satisfying $[a,\ada]=1$) and $\tir =
2/\sqrt{2(2\pi z^2)}$, the Hamiltonian matrix
$M_2 = U M_1 U^{-1}$ becomes
\eqn\nnmatfrm{M_2 = 4\pi z^2 R
\pmatrix{\ada a + \tir^2-1 & (\ada)^2 & \tir \ada \cr
a^2 & a\ada + \tir^2 +1 & -\tir a \cr \tir a & -\tir \ada & a\ada+\ada a \cr}.} 
Next, let us look for eigenvectors of this matrix. It is natural to
define a basis of harmonic oscillator eigenfunctions $L_n(\sigma)$ satisfying
$aL_n(\sigma) = \sqrt{n} L_{n-1}(\sigma)$ and $\ada L_n(\sigma) = \sqrt{n+1}
L_{n+1}(\sigma)$. The eigenvectors are then
of the form $(\alpha
L_n(\sigma), \beta L_{n-2}(\sigma), \gamma L_{n-1}(\sigma))$.
Letting $M_2$ act on a vector of this form, we can easily see that
it is transformed into a vector of the same type, with the matrix
$\tilde{M_2}$ acting on $(\alpha,\beta,\gamma)$, where

\eqn\tildemtwo{\tilde{M_2} = 4\pi R z^2 \pmatrix{\tir^2+n-1 & \sqrt{n(n-1)} &
\tir \sqrt{n} \cr \sqrt{n(n-1)} & \tir^2+n & -\tir \sqrt{n-1} \cr 
\tir \sqrt{n} &
-\tir \sqrt{n-1} & 2n-1}.}

Thus, all that is left to do is to find eigenvectors of this matrix,
and fortunately this is easy to do : there is one eigenvector $v_3 =
(-\sqrt{n},\sqrt{n-1},\tir)$ with eigenvalue $0$, and two eigenvectors
with eigenvalue $4\pi z^2 R(\tir^2+2n-1)=R(r^2+4\pi z^2(2n+1))$ 
which we can choose to be
$v_1=(\tir,0,\sqrt{n})$ and
$v_2=(\sqrt{n(n-1)},n+\tir^2,-\tir \sqrt{n-1})$. Note that the eigenfunction
corresponding to $v_1$ exists for any $n \ge 0$, while the
eigenfunction corresponding to $v_2$ exists only for $n \ge 2$.
The zero eigenvalue corresponds to a gauge transformation, which can
be ignored.

For the 6 other bosonic variables $A_\mu$, equation \baca\ immediately
implies that the eigenfunctions are again harmonic oscillator
eigenfunctions $L_n(\sigma)$, with eigenvalues (in the same 
normalization as above) $R(r^2+4\pi z^2(2n+1))$ for any $n \ge 0$.
These eigenvalues are (up to a factor $R$) the frequencies squared
of the
harmonic oscillators corresponding to the off-diagonal modes.

\subsec{Fermionic off-diagonal quadratic terms}

The term in the Hamiltonian \hamilton\ that gives a mass to the off-diagonal
fermionic modes is
\eqn\fermham{R\tr(\bt \gamma_i [\theta,Y_0^i]).}
Denoting by $\theta$ also the upper-right off-diagonal part of
$\theta$, we find for it a term of the form
\eqn\fermtr{R\tr(\bt \{ \gamma_1 (\theta Q_1 - Q_2 \theta) +
\gamma_2 (\theta P_1 + P_2 \theta) + \gamma_3 (-r\theta) \}).}
Translating into the field theory this becomes
\eqn\fermlag{R\int \bt \{ \gamma_1 (Q_1 - Q_2) + \gamma_2 (-P_1
+ P_2) - \gamma_3 r \} \theta.}

Now, it is easy to see that the mass matrix squared is simply (in the
same notations and normalizations as we used for the bosonic sector)
$M_f^2 = R^2(2(Q^2 + P^2) + r^2 -
2\gamma_1 \gamma_2 [Q,P])$. The eigenvalues of
$\gamma_1 \gamma_2$ are $\pm i$, and, thus, we find that
the eigenvalues of $M_f^2$ (with the usual eigenfunctions
$L_n(\sigma)$ defined above) are
$R^2(r^2 + 4\pi z^2(2n + 2))$ and $R^2(r^2 + 4\pi z^2(2n))$ for
$n=0,1,\cdots$. Since half of the fermions may be viewed as creation
operators and half as annihilation operators, we have $4$ states with
each eigenvalue.

Note that for a pair of membranes, the terms we were discussing in the
Hamiltonian (both for the bosons and for the fermions)
would depend only on $Q_1-Q_2$ and on $P_1+P_2$, which
commute with each other. Thus, all the mass shifts arising from the
non-commutativity would vanish, and the masses of all the bosonic and
fermionic off-diagonal modes would be the same, as required by the
unbroken supersymmetry in this case. 


\subsec{The membrane-anti-membrane potential}

Now we have all the information we need for computing the zero-point
energy of the membrane-anti-membrane configuration. The variable
$\sigma_+ \sim \sigma_1+\sigma_2$ 
decoupled from all our calculations, so that it just
corresponds to an (order $N$) degeneracy for all the modes we
found. The zero-point energy of each mode is the square root of the
matrices we discussed above (which correspond to $\omega^2$). Since
our variables are complex, the zero-point energy is the sum of the
frequencies we computed above (and not of half the frequencies). Joining
all of our results, we find that
the formula for the total zero-point energy is
\eqn\pot{\eqalign{V(r) = R
&\sum_{n=0}^{\infty} (6\sqrt{r^2+4\pi z^2(2n+1)} - 4\sqrt{r^2+4\pi z^2(2n)}
- 4\sqrt{r^2+4\pi z^2(2n+2)}) + \cr R&\sum_{n=0}^{\infty} 
\sqrt{r^2+4\pi z^2(2n-1)} + R
\sum_{n=2}^{\infty} \sqrt{r^2+4\pi z^2(2n-1)} \cr}}
or
\eqn\npot{\eqalign{V(r) = 
R \sum_{n=1}^{\infty} ( & \sqrt{r^2+4\pi z^2(2n+1)} - 4\sqrt{r^2+4\pi z^2(2n)}
+ 6\sqrt{r^2+4\pi z^2(2n-1)} - \cr & 4\sqrt{r^2+4\pi z^2(2n-2)} + 
\sqrt{r^2+4\pi z^2(2n-3)}) \equiv
\sum_{n=1}^{\infty} V(r,n). \cr }}
Although the sum of each term separately obviously does not converge,
$V(r,n)$ behaves for large $n$ like $n^{-7/2}$,
so the sum is well-defined. The fact that the coefficients in \npot\
are given by $(1,-4,6,-4,1)$, which are binomial coefficients,
guarantees the vanishing of the first four terms in the expansion of
$V(r,n)$ either in large $n$ or in large $r^2$.
In fact, we can perform the summation in \npot\ using the
Euler-Maclaurin formula
\eqn\eulermac{\sum_{k=1}^{\infty} F(k) = \int_0^\infty F(k) dk -
{1\over 2} F(0) - {1\over 12} F'(0) + {1\over 720} F'''(0) + \cdots}
which is valid for functions $F$ who vanish (with all their
derivatives) at infinity. Plugging in $F(k) = V(r,k)$, and restoring
the order $N$ degeneracy from the dependence on $x_+$, we find
that the integral behaves for large $r$ like $(-{3\over 16} (4\pi z^2)^3
R N r^{-5} +
O(r^{-7}))$, while the other terms vanish at least as fast as $r^{-7}$,
and in fact we can show that for large $r$ (compared to $z$)
\eqn\largepot{V(r) = -{{3 (4\pi z^2)^3 R N}\over {16 r^5}} + O({1\over
r^9}).}
This is the one-loop potential generated for $r$ in the quantum
mechanics, and it should correspond to the
potential for infinite membranes in eleven
dimensions, after integration over the separation of the membranes in
the longitudinal direction (since we are working with states of
definite longitudinal momentum).

Let us compare the potential we found with the supergravity result for
the long-range potential between a membrane and an anti-membrane. We
wish to find the correction to the rest mass of the system per
unit area. 
In
supergravity, the long-range gravitational force between two membranes
exactly cancels the long-range force arising from the 3-form
field. Thus, for a membrane-anti-membrane pair, the long-range
potential is exactly twice the gravitational potential. The metric
corresponding to a supermembrane solution of 11D supergravity
\duffstelle\ is (see, e.g., \duff)
\eqn\metric{g_{00} = (1 + {{2\kappa_{11}^2 T_M} \over {\Omega_7
r^6}})^{-2/3}}
where $\kappa_{11}$ is the 11D gravitational constant, $T_M$ is the
membrane tension, and $\Omega_7 = \pi^4/3$ is the volume of a
7-sphere of unit radius. $\kappa_{11}$ is related to the membrane
tension by $\kappa_{11}^2 T_M^3 = 2\pi^2$, and in the Newtonian limit
$g_{00} \sim 1 - 2V_g^{11}(r)$ where $V_g^{11}(r)$ is the gravitational
potential, so that we find in eleven dimensions $V_g^{11}(r) = -4 / \pi^2
T_M^2 r^6$. The membrane-anti-membrane potential per unit area is this
potential multiplied by the membrane tension, and restoring the factor
of 2 we find that
the total 11D potential is $V^{11}(r) = -8 / \pi^2
T_M r^6$. 
As we mentioned earlier, in their rest frame the membranes 
live on a compact circle of radius $R^{11}_{rest} = 1 / 2\pi z^2$. 
To get the M(atrix) theory potential which is in the lightcone frame, 
we should average over
the longitudinal direction, while adding images of the membranes due
to the periodicity requirement :
\eqn\relation{V^{LC}(r) = {1\over {2\pi R^{11}_{rest}}}
\int_{-\infty}^{\infty} dx_{11} V^{11}(\sqrt{r^2+{x^2_{11}}}) =
-{3 \over \pi (2\pi R^{11}_{rest}) T_M r^5}.}
As discussed in section 2, in the M(atrix) theory 
the membrane tension is given by $T_M=1/2\pi$, and plugging in the
value of $R^{11}_{rest}$ we find
$V^{LC}(r) = -{{6z^2} / r^5}$. 
Since the area of the membrane is $(2\pi)^2 N z^2$, we find that
the rest energy of this
configuration behaves as 
\eqn\resten{m \sim (2\pi)^2 N z^2 (2 \cdot {1\over {2\pi}} -
{{6z^2} \over r^5}).} 
Thus, we expect the leading order (in $1/r$) correction
to the M(atrix) theory calculation of $m^2/2P_-$ to be :
\eqn\result{2 \cdot ((2\pi)^2 N z^2)^2 \cdot 2 \cdot {1\over {2\pi}} 
\cdot {{-6 z^2}
\over r^5} / (2 \cdot 2N/R) = -{{48 \pi^3 N R z^6} \over r^5}.}
This is identical to what we found in \largepot, up to a factor of
$4$. We are not sure if this factor arises from a difference between
our conventions and the supergravity conventions, or if it arises from
the assumptions we made as to the
degeneracy of each state in the large $N$ limit. To the level of
precision we have been working in we
do not have good control over numerical factors arising from this
degeneracy.

The decay of the potential as $r^{-5}$, necessary for locality in eleven
dimensions, depended on many cancelations from the quantum mechanics
point of view. Naively, we would expect $V(r)$ in a theory which has
the same number of bosons and fermions to behave as $r$ for large
$r$. However, the fact that for theories with spontaneously broken
SUSY (as is the case in our computation) the sum of the masses squared
of the bosons still equals that of the fermions, ensures that $V(r)$
should decay at least as $1/r$. And indeed, this is the behavior we
would have found for the same computation in a different number of
transverse dimensions, since our
cancelations of the $1/r$ and $1/r^3$ terms
depended on $(1,4,6,4,1)$ being binomial coefficients. For instance,
in the analogous seven dimensional theory, the coefficients in \npot\
would have been $(1,2,2,2,1)$, and we would have found a potential
behaving like $1/r$ (which is, by the way, the expected potential for
membranes in this theory).
Presumably, the fact
that in our case we get a decay like $1/r^5$ is related to the larger
amount of SUSY that exists in the eleven dimensional 
theory, but the exact way in
which this works is still unclear to us. For instance if, in theories
with $16$ supersymmetry generators, there are
also sum rules on the fourth and sixth powers of the masses, that
would explain our results, but we do not know of the existence of such
sum rules.

Another way in which one may try to understand 
the cancelation to this order, which
was suggested to us by M. Douglas and S. Shenker, is the
following. The leading correction to the 0-brane action\dbranethree, 
which gives the
desired graviton-graviton scattering, is given by a term
of order $v^4/r^7$, whose $D=10$ SYM origin is an abelian $F_{\mu
\nu}^4$ term. The non-abelian completion of this term, if such a term
exists, has the commutator in it. If we can treat the
membrane-anti-membrane as soft breaking of 
SUSY, the term $v^4/r^7$ will be
replaced by $z^8/r^7$. Integrating over two coordinates of the
membrane this becomes $z^6/r^5$, as we found above. 

\subsec{Membrane-anti-membrane annihilation}

For $r^2=4\pi z^2$, one of the frequencies we computed vanishes, while for
$r^2<4\pi z^2$
the field theory develops a tachyon, signaling an instability in the
configuration. Note that since $R^{11}_{rest} = 1 / 2\pi z^2$, 
this distance is exactly proportional to the string scale in our
formalism, where a similar instability may be found for a
D-brane-anti-D-brane configuration. 
Of course, for small values of the frequencies, our one-loop
approximation is no longer valid, and we can no longer neglect the
excitations of the off-diagonal modes. However, when the string scale
is large compared to the Planck scale (i.e. the radius of the eleventh
dimension is small), the other off-diagonal modes will still have high
frequencies when the tachyon develops, and it is legitimate to ignore
them and assume that only the tachyonic mode is excited.

The wave function corresponding to
the ground state of this tachyonic mode is of the form $A_1 \sim
e^{-{1\over 2}(\sigma_1-\sigma_2)^2}$, 
$A_2 \sim ie^{-{1\over 2}(\sigma_1-\sigma_2)^2}$,
and this can be multiplied by any function of
$\sigma_1+\sigma_2$. Since we interpreted the $\sigma_i$ as positions
along one of the coordinates of the membrane (recall that they came
from the $Q_i$), such an excitation is localized at the same place in
both membranes. Note that in momentum space, which we identify with
the second coordinate of the membrane, the tachyonic mode is
also proportional to $e^{-{1\over 2}(\del_1-\del_2)^2}$, so the
excitation is in fact localized in both membrane coordinates. Adding
an arbitrary function of $\sigma_1+\sigma_2$ just corresponds to a
superposition of many such local excitations along the surface of the
membranes. When the tachyon condenses, the fields corresponding to the
distance between the two membranes become massive, so that the
membranes tend to join together, and presumably eventually annihilate
into gravitons. Presumably, this may be interpreted as the
condensation of a string between the two membranes, which in M theory
is identified with a ``wormhole'' configuration connecting the
membrane and the anti-membrane \asy. The condensation of such strings
``eats up'' the surface of the membranes, and they annihilate into
gravitons. It is less clear what happens in the eleven dimensional
limit, when the string scale (where the tachyonic mode arises) is
much smaller than the Planck scale. In this case we cannot trust our
approximations, and different methods should apparently be used to
analyze the annihilation.

\newsec{Membranes at finite longitudinal velocities}

 From the point of view of supergravity, gravitons are considerably
simpler then membranes. In M(atrix) theory, 
the situation is reversed, in the sense
that we have a full description of the infinite membranes whereas the
graviton wave functions remain elusive. Realizing this we can now
re-examine the issue of Lorentz invariance, which we could not have
done in the supergravity multiplet sector due to lack of knowledge of
the bound state wave function.  In the following we will take the
first step towards showing Lorentz invariance in the membrane
sector. We will show that the M(atrix) model computation reproduces
the supergravity potential between a pair of infinite membranes (in
the kinematical setup discussed above) when they are moving with a
small relative longitudinal velocity. This implies that the M(atrix)
model will reproduce processes with longitudinal momentum transfer to
the first sub-leading contribution to the moduli space approximation.

Let us first discuss the kinematical setup and the supergravity
predictions. We will now take a pair of membranes satisfying
$[P_i,Q_i]=-2\pi iz_i^2$, $i=1,2$. Transforming to the rest frame of
the first membrane, the velocity of the second membrane is proportional to
$(z_2^2-z_1^2) / z^2$ (to leading order in $z_1-z_2$). We work only
to leading order in $z_1-z_2$, and in quantities that do not depend on
the difference we will denote both $z_1$ and $z_2$ by $z$. In this
section we will not put in all of the numerical factors, and will
check only the dependence of the results on $N,R,r$ and the $z_i$'s.

The supergravity result for the potential 
in the approximate rest frame (where the
velocities are small), in the eleven dimensional theory,
is $Vol\times {v^4 / r^6}$, where $Vol$ is the area of the two 
membranes. Since we are dealing with a
configuration that in the rest frame is compactified on a circle of
radius ${1 / z^2}$, this is modified (upon averaging over the
longitudinal direction) to $Vol\times {z^2 v^4 /
r^5}$. Recalling the $z$ dependence of the area, we obtain that the
potential in the rest frame is ${Nz^4v^4 / r^5}$. In the Hamiltonian
we therefore expect a correction ${1\over {N/R}}{{Nz^2 \cdot Nz^4 
v^4}\over
r^5}={1\over {N/R}}N^2{(z_1^2-z_2^2)^4\over z^2}{1\over r^5}$

Let us perform now the M(atrix) model calculation. We implement this
configuration by taking $Q_i=z_i\sigma_i,\
P_i=-iz_i\partial_{\sigma_i}$, and we use the  method described
above to write the Hamiltonian of the off-diagonal terms as a simple 2+1
field theory. As we will see shortly, the numerical details of the
phase-space matching are more subtle than in the membrane
anti-membrane case (since the field
theory changes qualitatively when $z_1=z_2$), but it is still easy to
extract the dependence of the answer on $z,N,R$ and $r$.

Let us take, for example, the Hamiltonian for coordinates transverse
to the brane. We obtain that the potential term is
\eqn\pttr{\int d\sigma_1d\sigma_2
A_\mu^* (r^2+(Q_1-Q_2)^2+(P_1+P_2)^2)A_\mu.}
Changing to variables
$\sigma_+={z_1\sigma_1+z_2\sigma_2\over\sqrt{z_1^2+z_2^2}}$ and
$\sigma_-={z_2\sigma_1-z_1\sigma_2\over\sqrt{z_1^2+z_2^2}}$,
we obtain that the Hamiltonian is (to leading order in $z_1^2-z_2^2$) 
\eqn\ptb{\int d\sigma_+ d\sigma_- A^*_\mu(r^2+{(z_1^2-z_2^2)^2\over
z^2}(\sigma_+ +{2z_1 z_2 \over {z_1^2-z_2^2}}\sigma_-)^2
-z^2\partial_+^2)A_\mu.}

The energy
levels of this field theory, due to the harmonic oscillator in the
$\sigma_+$ direction, are of the form $r^2+(z_1^2-z_2^2)(2n+1)$. 
Similar results pertain to all
the other bosonic and fermionic modes, and we find that the sum over
frequencies (performed in the same way as in the previous section) 
is proportional to $(z_1^2-z_2^2)^3 / r^5$. 

However, in this case we need to be more careful in going to the field
theory limit. We should check more carefully what is the upper value
of $n$ that we should take (for finite $N$), and, correspondingly,
what is the degeneracy that arises from the integration over
$\sigma_-$ (since the total number of modes is $N^2$).
We can determine the highest allowed value of $n$ (for finite $N$) as
follows. The size of the membranes is $[-z\sqrt{N},z\sqrt{N}]$ in
each direction, so the most massive state should have a mass of order
$r^2+z^2N$. On the other hand, the states we find for each $n$
have energies of order $r^2+(z_1^2-z_2^2)(2n+1)$. 
Equating these two energies, we conclude that the highest allowed
value of $n$ should be $N{z^2 \over {z_1^2-z_2^2}}$, and,
correspondingly, the degeneracy of each level is of order
$N{{z_1^2-z_2^2}\over z^2}$.
Thus, we find that the M(atrix) theory result is
${1\over{N / R}}N^2{(z_1^2-z_2^2)^4\over z^2}{1\over r^5}$, which is
the same as the supergravity result (up to numerical factors).

Note that if we (not necessarily justifiably) extend our calculations
to small values of $r$, we find now a tachyon appearing for $r \sim
\sqrt{z_1^2-z_2^2} \sim \sqrt{v} z$ and not at the string scale $r
\sim z$. The scale $r \sim \sqrt{v}$ (in string units) 
is known to be the
relevant ``interaction'' scale for moving D-branes in string theory
\refs{\bachas,\dbranethree}, and we see that it appears also in
M(atrix) theory. We can easily generalize the calculation above to
the case of a membrane and an anti-membrane moving at a relative
longitudinal velocity, and we find the expected $v^2/r^5$ correction
in this case.

\newsec{Membrane-0-brane dynamics}

The computation for a membrane and a 0-brane (or a graviton state) 
is exactly the same as that of the
membrane-anti-membrane case, if we take the second block to be of size
$1\times 1$ (instead of $N\times N$), and set $Q_2=P_2=0$. Thus, our
field theory will just be $1+1$ dimensional, and we will find simple
mass matrices for all the fields. For the bosonic transverse modes the
mass matrix squared is $R^2(r^2+Q^2+P^2) \to R^2(r^2+2\pi z^2(2n+1))$, for
the fermions it
is $R^2(r^2+Q^2+P^2+\gamma_1\gamma_2[Q,P]) \to R^2(r^2+2\pi z^2(2n+1\pm
1))$, and for the
longitudinal bosons we again find the same split as before. For $N_1$
0-branes each mode has an additional degeneracy of $N_1$ (we assume
$N_1 \ll N$, since we take the $N \to \infty$ limit first).
The total
potential thus comes out to be, by a similar computation to
the membrane-anti-membrane case,
\eqn\zerotwopot{V(r) = -{{3 (2\pi z^2)^3 N_1 R}\over{16 r^5}} + O({1\over
r^9}).}

Again, we should compare this with the expected supergravity
result. For this we again transform to the rest frame of the
membrane. In the light cone frame, the 0-branes had $P^+ = N_1 /
R$. In the rest frame, this is transformed to $P^+ = N_1 \sqrt{2} \pi
z^2$, and this is also (up to a $\sqrt{2}$ factor)
the energy of the 0-branes in this frame. The result we expect to find
is the same as the result in the membrane-anti-membrane case,
but with
the energy of the 0-brane replacing the mass of the membrane (up to a
numerical constant).
This is because from the 10D perspective the energy of
the 0-brane is simply its mass, and it is also at rest, and both the
0-brane-2-brane and the 2-brane-anti-2-brane 
potentials are proportional to the
corresponding gravitational potentials (but with a different constant
of proportionality).
The ratio of the 10D masses is ${{N_1 \pi z^2} / {2\pi
z^2 N}} = N_1 / 2N$, so we expect the 0-brane result to be $N_1 / N$
times the membrane-anti-membrane result \largepot, up to numerical
factors, and this is indeed what we find in \zerotwopot.

As in the membrane-anti-membrane case, we find a tachyonic mode at the
string scale. A similar mode exists for a D-0-brane D-2-brane
configuration in type IIA string theory. The wave function of the
tachyon now looks like $e^{-{1\over 2}x^2}$ where $x$ is the position
of the 0-brane, so it is localized (in the membrane worldvolume) near
the position of the 0-brane. The condensation of this tachyon makes
the modes corresponding to the separation between the 0-brane and the
membrane massive, so that they would tend to join
together. Presumably, the 0-brane is swallowed into the membrane
worldvolume, corresponding to a boost of the membrane from the eleven
dimensional
point of view, or to a gauge field inside the D-2-brane from the ten
dimensional
point of view. The corresponding 
bound states of the D-0-brane and the D-2-brane in type IIA string
theory were recently discussed in \bound. As in section 3, it is less
clear what happens in the eleven dimensional limit, since our
approximations are not necessarily valid then.

\newsec{Conclusions and open questions}

In this paper we computed the long range potentials associated with
various configurations involving infinite membrane in M(atrix) theory,
and compared the results with the known supergravity potentials. In
all three cases we discussed (the membrane-anti-membrane potential,
the potential between membranes at relative longitudinal velocities
and the potential between a membrane and a graviton), we found that
the M(atrix) theory computation agrees with the supergravity result,
up to numerical factors. We believe this provides more evidence that
M(atrix) theory indeed describes M theory membranes correctly, and
that it is Lorentz invariant (though, obviously, more tests are
needed to establish exact Lorentz invariance).

Our computations in the quantum mechanics involved only the zero point
energies of the off-diagonal matrices, ignoring their
fluctuations. This is valid when the distance between the membranes
(or gravitons) in our computations is large, so this is the only
regime where we can trust our calculations. When the distance is of
the order of the string scale, we find that in the
membrane-anti-membrane and graviton-membrane
configurations a tachyon (a mode with negative
$\omega^2$) appears. We can trust this result only if the string scale
is larger than the eleven dimensional Planck scale, where we expect
additional corrections could arise, and then our tachyons presumably
correspond exactly to the corresponding tachyons arising from open
strings between D-branes in the type IIA theory (which is then weakly
coupled). It is less clear what we can say about the opposite limit,
in which the eleventh dimension is large, and the string scale is much
smaller than the Planck scale. This is an issue that deserves further
investigation. 

In our computations we worked only with infinite flat
membranes. It would be interesting to find a direct connection between
these configurations and the finite, closed membranes which also exist
(as long-lived semi-classical configurations) in M(atrix) theory. Upon
compactification, the infinite membrane solutions, if wrapped around
compact directions, become finite energy solutions, and the problems
we encounter due to the infinite energy and size of the membranes
should disappear. Thus, it should be interesting to perform similar
computations to ours with wrapped membranes in compactified M(atrix)
theory.

\centerline{\bf Acknowledgements}

We would like to thank Tom Banks, Michael Douglas, Nati Seiberg and
Steve Shenker for many useful discussions. This work was supported in
part by DOE grant DE-FG02-96ER40559.

\listrefs

\end